\documentclass{JHEP3}
\usepackage{amsmath,amsfonts,bbm,euscript}
\usepackage{cite}
\usepackage{epsfig}

\newcommand{\mb}[1]{\mathbf{#1}}

\renewcommand{\bar}[1]{
  \overline{#1}
}
\def\Db{$\bar{\text{D}}$}
\def\Td{T^\dagger}
\def\ba{\begin{eqnarray}}
\def\ea{\end{eqnarray}}
\def\be{\begin{equation}}
\def\ee{\end{equation}}     
\def\bas{\begin{eqnarray*}}
\def\eas{\end{eqnarray*}}
\def\Str{\text{Tr}^*\;}
\def\Tr{\text{Tr}\;}
\def\Str{\text{Tr}^*\;}
\def\det{\text{det}}
\def\ap{{\alpha^{\prime}}}
\def\hf{\frac12}
\DeclareMathOperator{\F}{\EuScript{F}}
\def\G{G}
\def\X{\mathcal{X}}
\def\Y{\mathcal{Y}}
\def\DD{\text{D}\bar{\text{D}}}
\def\DDD{\text{DD}\bar{\text{D}}}

\def\b{\bar}
\def\p{\partial}

\def\FFFF{\F\left(\X^+ + \sqrt{\Y^+}\right)\F\left(\X^+ - \sqrt{\Y^+}\right)\F\left(\X^- + \sqrt{\Y^-}\right)\F\left(\X^- - \sqrt{\Y^-}\right)}
\def\la{2\pi\ap}

\title{Adding a Brane to the Brane-Anti-Brane Action in BSFT}

\author{Nicholas T. Jones,  Louis Leblond, and  S.-H. Henry Tye\\
  Laboratory for Elementary-Particle Physics,
 Cornell University, Ithaca, NY 14853.\\
  E-mail: \email{nick.jones@cornell.edu},
\email{lleblond@mail.lepp.cornell.edu}
  \email{tye@mail.lepp.cornell.edu}}

\abstract{We attempt to generalize the effective action for the 
D-brane-anti-D-brane system obtained from boundary superstring 
field theory (BSFT) by adding an extra D-brane to it to obtain a
covariantized action for 2 D-branes and 1 anti-D-brane.
We discuss the approximations made to obtain the effective 
action in closed form.
Among other properties, this effective action admits solitonic 
solutions of codimension 2 (vortices) when one of the D-brane 
is far separated from the brane-anti-brane pair. }


\begin{document}

\section{Introduction}

The study of non-BPS brane system has revealed a lot of 
interesting properties of the superstring theory.  
In particular, the tachyon rolling \cite{Sen:2002nu,Sen:2002in} 
in the non-BPS-D-brane 
and the D-brane-anti-D-brane ($\DD$) pair allows the study of 
string dynamics. A particularly powerful approach is the 
effective actions\cite{Kraus:2000nj,Takayanagi:2000rz,Jones:2002si}
derived from boundary superstring field theory (BSFT)
\cite{Witten:1992qy,Gerasimov:2000zp,Kutasov:2000aq}.
It is natural to generalize the approach to more complicated systems.
In this paper, we attempt to derive the effective action for the
2 D-branes and 1 anti-D-brane ($\DDD$) system in BSFT. There are
a number of motivations to study this system. The system is more 
complicated and we like to see how that translates into the 
formalism and the formulation of the effective action. We also 
like to use the condensation of the tachyon in this system 
to produce an action 
for other systems, for example, that of a Dp-brane plus a 
D(p-2)-brane system, where the D(p-2)-brane is a codimension-2 
topological defect resulting from the tachyon rolling. On the more
phenomenological side, we see that the decay of the $\DD$ pair in 
the $\DDD$ system produces both open string modes as well as closed 
string modes, a feature that is absent in the $\DD$ system 
since there is no open strings modes left after the annihilation
of the branes. This property is particularly important in brane 
inflation, where we like to see most of the energy released 
from brane collisions to go to reheating the universe, that is,
to open string modes instead of closed string modes. This is 
required by the big bang nucleosynthesis.

Apriori, it is clear that the $\DDD$ system involves non-abelian 
gauge theory and any effective action one can obtain in closed 
form is a poorer approximation to the actual theory than that 
for the $\DD$ action. Fortunately, there are still many 
interesting features maintained in a simplified closed form
$\DDD$ effective action.

The paper is organized as follows. In \S2, we briefly review the 
BSFT derivation of the $\DD$ effective action. 
In \S3, we present the covariantized $\DDD$ effective action.
As a check, we show how our new action reduces correctly to 
the $\DD$ effective action when one of the D-brane is moved to 
$\infty$. We leave the somewhat lengthy determination 
of invariants to the appendix. 
In \S4, we construct the effective action 
for two D$p$-branes and one $\b{\text{D}}p$-brane by T-dualizing
the $\DD$ effective action. This allow us to physically 
justify our approximate effective action. Finally, we study the solitonic
solutions of our action in \S5.
\S6 is the conclusion.

\section{Review of BSFT and the $\DD$ system}

We review the $\DD$ effective action from BSFT
derived in Ref\cite{Kraus:2000nj,Takayanagi:2000rz}. 
We restrict our attention to D9-branes in type IIB theory.
BSFT essentially extends the sigma-model
approach to string theory\cite{Tseytlin:1989rr}, in that (under
certain conditions \cite{Witten:1992qy,Gerasimov:2000zp}) the disc
world-sheet partition function with appropriate boundary insertions
gives the classical spacetime action. This framework for the bosonic
BSFT was extended to the open superstring in \cite{Kutasov:2000aq} and
formally justified in \cite{Marino:2001qc,Niarchos:2001si}.  In the NS sector the
spacetime action is
\begin{align}\label{definitionS}
  S_{\text{spacetime}} &= -\int \mathcal DX\mathcal D\psi
    \mathcal D\tilde\psi\;e^{-S_\Sigma-S_{\partial\Sigma}}.
\end{align}
where $\Sigma$ is the worldsheet disc and $\partial\Sigma$ is its
boundary.  The worldsheet bulk disc action is the usual one
\begin{align*}
  S_\Sigma &= \frac1{2\pi\ap}\int d^2z\;
  \partial X^\mu\bar\partial X_\mu 
  + \frac1{4\pi}\int d^2z\left(\psi^\mu\bar\partial\psi_\mu 
  + \tilde\psi^\mu\partial\tilde\psi_\mu\right)\\
  &= \hf\sum_{n=1}^\infty nX_{-n}^\mu X_{n\;\mu} + 
  i\sum_{r=\hf}^\infty\psi_{-r}^\mu\psi_{r\;\mu},
\end{align*}
after expanding the fields in the standard modes.  To reproduce the
Dirac-Born-Infeld (DBI) action for a single brane, the appropriate
boundary insertion is the boundary pullback of the $U(1)$ gauge
superfield to which the open string ends couple; for the $N$ brane $M$
anti-brane system, the string ends couple to the superconnection
\cite{Quillen,Witten:1998cd}, hence the boundary insertion should be
\begin{align}\label{BoundaryInsertion}
  e^{-S_{\partial\Sigma}} &= \Tr\hat P\exp\left[\int d\tau d\theta 
    \mathcal M(\mb X)\right],&
  \mathcal M(\mb X) &= \left(\begin{array}{cc}
    iA^1_\mu(\mb X)D\mb X^\mu&\sqrt\ap T^\dagger(\mb X)\\
    \sqrt\ap T(\mb X)&iA^2_\mu(\mb X)D\mb X^\mu
  \end{array}\right)
\end{align}
where the insertion must be supersymmetrically path ordered to
preserve supersymmetry and gauge invariance.  $A^{1,2}$ are the $U(N)$
and $U(M)$ connections, and $T$ is the tachyon matrix transforming in
the $(N,\bar M)$ representation of $U(N)\times U(M)$. The lowest 
component of
$\mathcal M$ is proportional to the superconnection.  To proceed, it
is simplest to perform the path-ordered trace by introducing boundary
fermion superfields \cite{Marcus:1987cm}; we refer the reader to
\cite{Kraus:2000nj} for details.  The insertion
(\ref{BoundaryInsertion}) can then be simplified to be
\begin{align}\label{NMinsertion}
  \Tr P\exp\left[i\ap\int d\tau
    \left(\begin{array}{cc}
      F^1_{\mu\nu}\psi^\mu\psi^\nu+iT^\dagger T 
      + \frac1\ap A^1_\mu\dot X^\mu&
      -iD_\mu T^\dagger\psi^\mu\\
      -iD_\mu T\psi^\mu&
      F^2_{\mu\nu}\psi^\mu\psi^\nu+iTT^\dagger + 
      \frac1\ap A^2_\mu\dot X^\mu
    \end{array}\right)\right],
\end{align}
where the tachyon covariant derivatives are
\begin{align}
\label{Tcderiv}
  D_{\mu}T = \partial_{\mu} T + iA^1_{\mu}T-iTA^2_{\mu}
=\partial_{\mu} T + iA^-_{\mu}T
\end{align}
This expression reproduces the expected results when
the tachyon and its derivatives vanish: the only open string
excitations will be the gauge fields on the branes and the
anti-branes, for each of which the action is the standard DBI action.
For a single brane anti-brane pair, $N=M=1$, demanding that the
gauge field to which the tachyon couples vanishes, $A^-=0$,
the path-ordered trace can be performed using worldsheet
boundary fermions.  Writing $A^+=A^1+A^2$, we have 
\begin{align}\label{1DDinsertion}
  S_{\partial\Sigma} &= -\int d\tau\left[\ap T\bar T + 
      \ap^2(\psi^\mu\partial_\mu T)\frac1{\partial_\tau}
      (\psi^\nu\partial_\nu\bar T) + \frac i2\left(\dot X^\mu A^+_\mu
      +\hf\ap F^+_{\mu\nu}\psi^\mu\psi^\nu\right)\right].
\end{align}
\begin{align}
\frac{1}{\partial_\tau} f(\tau) = \int d\tau^{\prime} f(\tau^{\prime}) 
sgn(\tau -\tau^{\prime}) 
\end{align}
where $sgn(x)=1$ for $x>0$ and $=-1$ for $x<0$.
For linear tachyon profiles, spacetime
rotations allow us to write $T=u_1X^1+iu_2X^2$, and
(\ref{definitionS}) can be calculated, since the functional integrals
are all Gaussian.  The result when $A^+=0$
is derived in \cite{Kraus:2000nj,Takayanagi:2000rz}:
\begin{align}\label{NonCovAction}
  S_{\DD} = -2\tau_9\int d^{10}X_0\;\exp&\left[
    -2\pi\ap[(u_1X_0^1)^2+(u_2X_0^2)^2]\right]\F(4\pi\ap^2 u_1^2)
  \F(4\pi\ap^2 u_2^2).
\end{align}
where the function $\F(x)$ is given by\cite{Kutasov:2000aq} 
\begin{align}\label{Fdefinition}
  \F(x) = \frac{4^xx\Gamma(x)^2}{2\Gamma(2x)}
  = \frac{\sqrt{\pi}\Gamma(1+x)}{\Gamma(\hf+x)}.
\end{align}
\begin{align}  
\F(x) = \begin{cases}
    1 + (2\ln2)x + \mathcal O(x^2),&
    0<x\ll1,\\\label{Fseries}
    \sqrt{\pi x}  & x\gg1.
  \end{cases} 
\end{align}

The RR coupling of the D-branes can also be written down.
The bulk contribution to the
partition sum can be written as the wave-functional
\cite{Kraus:2000nj,Takayanagi:2000rz}
\begin{align*}
  \Psi^{RR}_{\text{bulk}} &\propto\exp\left[
    -\hf\sum_{n=1}^\infty nX_{-n}^\mu X_{n\;\mu} - 
  i\sum_{n=1}^\infty\psi_{-n}^\mu\psi_{n\;\mu}\right]C,\\
  C&=\sum_{\text{odd }p}\frac{(-i)^{\frac{9-p}2}}{(p+1)!}
  C_{\mu_0\cdots\mu_p}\psi_0^{\mu_0}\cdots\psi_0^{\mu_p}.
\end{align*}
The $\psi_0^\mu$ are the zero modes of the Ramond sector fermions, and
$C_{\mu_0\cdots\mu_p}$ are the even RR forms of IIB string theory.  The
normalization of $\Psi$ can be set later by demanding that the correct
brane charge is reproduced.  The Chern-Simon-like action is then defined
by
\begin{align*}
  S_{\text{RR}} &= \int\mathcal DX\mathcal D\psi\;\Psi^{RR}_{\text{bulk}}
  \Str P e^{-S_{\partial\Sigma}},
\end{align*}
in which the trace given by
\begin{align*}
  \Str O \equiv \text{Tr}\;\left[\left(
    \begin{array}{cc}\mathbbm{1}_{N\times N}&0\\
      0&-\mathbbm{1}_{M\times M}\end{array}\right)O\right]
\end{align*}
results from the periodicity of the worldsheet fermion superfield
which was necessary to implement to the supersymmetric path ordering.
By Witten's
argument \cite{Witten:1982df}, only the zero modes contribute to the
partition sum, giving \cite{Kennedy:1999nn,Kraus:2000nj,
Takayanagi:2000rz}
\begin{align}\label{generalCS}
  S_{\text{RR}} &= \tau_9g_s\int C\wedge\Str e^{2\pi\ap i\mathcal F},\\
  \mathcal F &= \left(\begin{array}{cc}
      F^1 + iT^\dagger T & -i(DT)^\dagger\\
      -iD T & F^2 + iTT^\dagger
    \end{array}\right)\nonumber
\end{align}
$\mathcal F$ is the curvature of the superconnection, and as usual,
the fermion zero modes form the basis for the dual vector space and
all forms above are written with $\psi_0^\mu \to dx^\mu$.  This
expression is exact\footnote{As discussed in \cite{Kraus:2000nj}, this
action is exact in $T$ and $A^\pm$ and their derivatives, but has
corrections for non-constant RR forms.}  and although it was derived
for $2^{m-1}$ brane anti-brane pairs in \cite{Kraus:2000nj,
Takayanagi:2000rz} it appears to have the correct properties for the
general $N$ brane $M$ anti-brane case.

\subsection{Covariant action}

The above $\DD$ action may be further improved by making it
manifestly covariant \cite{Jones:2002si}.
By requiring covariance of the action, it was possible to 
generalize the previous action to any tachyon profile. 
The main step in this improvement was to recognize that there is
two independent U(1) and Lorentz invariants made of first 
derivatives in the tachyon fields.\footnote{The metric is understood 
as being the open string metric $G^{\mu\nu} = 
\left(\frac{1}{G-B}\right)^{(\mu\nu)}$.  Here, the antisymmetric
part is set to zero.} 
\begin{align*}
  \X &\equiv 2\pi\ap^2g^{\mu\nu}\partial_\mu T\partial_\nu\bar T,&
  \Y &\equiv\left(2\pi\ap^2\right)^2
  \Big(g^{\mu\nu}\partial_\mu T\partial_\nu T\Big)
  \Big(g^{\alpha\beta}\partial_\alpha\bar T\partial_\beta\bar T\Big),
\end{align*}
With the normalizations chosen for convenience.  For the linear
profile $T=u_1x_1+iu_2x_2$, the only translation invariant way to
reexpress $u_{1,2}$ is as $u_{1,2} = \partial_{1,2} T^{1,2}$; then
with $g^{\mu\nu}=\eta^{\mu\nu}$ we can calculate $\X$ and $\Y$,
\begin{align*}
  \X&=2\pi\ap^2(u_1^2 +u_2^2),\\
  \Y&=\left(2\pi\ap^2\right)^2(u_1^2-u_2^2)^2,
\end{align*}
so the arguments of $\F$ in (\ref{NonCovAction}) can be written as
\begin{align*}
  4\pi\ap^2u_1^2 &= \X+\sqrt{\Y},\\
  4\pi\ap^2u_2^2 &= \X-\sqrt{\Y}.
\end{align*}
This provides a unique way to covariantize (\ref{NonCovAction}) as
\begin{align}\label{DDnoGaugeAction}
  S_{\DD} = -2\tau_9\int d^{10}x\sqrt{-g}\;&e^{-2\pi\ap T\bar T}
  \F(\X+\sqrt \Y)\F(\X-\sqrt \Y)
\end{align}
Using symmetry constraint, one can also restore the gauge field coupling
via the minimal coupling : 
$\partial_{\mu} T \to D_{\mu}T=\partial_{\mu}T +i A^-_{\mu}T$.

\section{Multiple Branes-Anti-Branes Action}

\subsection{N D9-branes and M \Db 9-branes}

Let us consider the generalization of the action for $\DD$ to 
that of $N$ D-branes and $M$ anti-D-branes.  We will
then point out where only approximative results are possible or 
where a perturbative expansion should be performed (the path 
integral is not gaussian).

For a system of $N$ D9-branes and $M$ \Db 9-branes (throughout we will
assume $N\ge M$), the tachyon is that set of fields which stretches
between the branes and the anti-branes; hence $T$ is an $N\times M$
matrix, which transforms in the bifundamental of the $U(N)\times U(M)$
gauge group of the system.  The potential for the tachyon fields can
be derived from the work of \cite{Kraus:2000nj, Takayanagi:2000rz},
who present the world-sheet sigma-model action for such systems, and
evaluate the action for the tachyon when $N=M=1$.  The tachyon
potential can be readily evaluated for general $N$ and $M$, beginning
with the world-sheet boundary insertion (\ref{BoundaryInsertion}).
Using the definition of supersymmetric path ordering as written in
\cite{Kraus:2000nj},
\begin{align}\nonumber
  \hat P\exp\left[\int d\hat\tau
    \mathcal M(\hat\tau)\right]
  &= \sum_{N=0}^\infty\int d\hat\tau_1\ldots d\hat\tau_N
  \Theta(\hat\tau_{12})\Theta(\hat\tau_{23})
  \ldots\Theta(\hat\tau_{N-1,N})
  \mathcal M(\hat\tau_1)\ldots\mathcal M(\hat\tau_N)\\\nonumber
  &= 1 + \int d\tau \left(\mathcal M_1(\tau) - \mathcal M_0^2(\tau)
  \right)+\ldots\\\label{SuPathOrder}
  &= P\exp\left[\int d\tau\left(
    \mathcal M_1(\tau) - \mathcal M_0^2(\tau)\right)\right]
\end{align}
in which $\hat\tau_{12} = \tau_1-\tau_2+\theta_1\theta_2$, $P$ in
the result is normal path ordering after integration over superspace,
and $\mathcal M_{0,1}$ are the parts of the matrix $\mathcal M$ which
are proportional to zero and one power of $\theta$.

For constant tachyons, and vanishing gauge fields,
\begin{align*}
  &\mathcal M_0 = \sqrt\ap\left(\begin{array}{cc}
    0&\Td\\ T&0
  \end{array}\right),& \mathcal M_1 = 0,
\end{align*}
and the path ordering is irrelevant, so the potential for a general
numbers of branes and anti-branes is simply 
\begin{align}\nonumber
  V(T,\Td) &= \tau_9\Tr\exp\left[-2\pi\ap
    \left(\begin{array}{cc}
        \Td T&0\\0&T\Td
      \end{array}\right)
  \right]\\
  &= \tau_9\left[(N-M) + 2\Tr e^{-2\pi\ap T\Td}\right].
  \label{Potential}
\end{align}
This potential clearly agrees with all the physics expected from Sen's
arguments \cite{Sen:1999mg}; when $T=0$, since the branes and
anti-branes are frozen at the unstable maximum, the potential is just
the tension of $(N+M)$ D9 branes.  Also, it is known that the $M$
anti-branes should decay with $M$ of the branes, leaving only $(N-M)$
D9-branes; at its stable minimum, $|T|\to\infty$, the potential takes
the value $\tau_9(N-M)$, which is the tension of the remaining branes.

For non-constant tachyon, the path integral is not gaussian and cannot be
performed in general.  Nevertheless we might be able to describe some 
parts of this system approximatively.  The calculations are quite 
involved for the generic case so, for the rest of this paper, we will 
focus on 2 D9-branes and 1 \Db 9-brane.

\subsection{Two D9-branes and 1 \Db 9-brane}

\EPSFIGURE{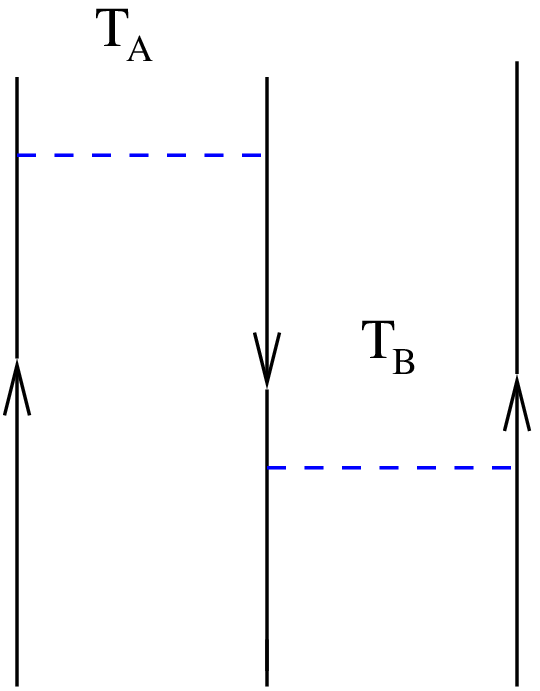}{Two D-branes (arrow up) and 1 anti-D-brane 
(arrow down) with two complex tachyons stretching between them.  
The branes are all sitting on top of each other.}
For the case of 2 D9-branes and 1 \Db 9-brane,
the boundary insertion is a $3 \times 3$ matrix.  
The boundary fermions method use in \cite{Kraus:2000nj} to perform the path
ordering is applicable for a 
matrix that can be expanded in terms of $SO(2m)$ generators. 
This is not the case here. 
One way to circumvent this problem is to extend the
matrix to a $4 \times 4$ matrix, set the extra tachyons to zero
and then use boundary fermions as before.  
We then get a non-gaussian path integral which
can only be done perturbatively.
Another way is to do the path ordering by hand term by term 
in the expansion series. This way turns out to be useful to see 
the behavior of the action and
to see what kind of approximation we will need to do.

We start with the supersymmetrically ordered expression (\ref{SuPathOrder})
which is valid for any boundary insertion of the form 
(\ref{BoundaryInsertion}).
We will set the gauge field and the Kalb-Ramond fields to 0 but 
we otherwise 
consider a general case of $2$ D9-branes and $1$ \Db9-brane. The boundary
insertion is then (after integrating over $\theta$):
\begin{align}
M_1(\tau) - M_0^2(\tau) &= M(\tau) = \ap \left( \begin{array}{cc}
             -T^\dagger T & \sqrt{\ap}\psi^\mu\p_\mu T^\dagger\\
              \sqrt{\ap}\psi^\mu\p_\mu T & -TT^\dagger\\
	      \end{array}\right)
\end{align}
\begin{align*}
T & = \left(\begin{array}{cc}
            T_A & T_B\\
	    \end{array}\right) \qquad
T^\dagger = \left(\begin{array}{c}
            \b T_A\\
	    \b T_B\\
	    \end{array}\right)\\
T_A &= T_1 + iT_2  \qquad T_B = T_3 + iT_4\\
\end{align*}
where $T_A$ and $T_B$ are the two complex fields between the two 
branes and the anti-brane as shown in figure 1.  
$T_1\dots T_4$ are 4 real tachyon fields.

The path ordering can be done by hand:
\begin{align}\label{Expansion}
e^{-S_{\p\Sigma}}& = \Tr\hat P\exp\left[\int d\tau 
    \mathcal M(\tau)\right]\\\nonumber
&=\tau_9 \Tr \sum_{N=0}^{\infty} \int d\tau_1\hdots d\tau_N
\Theta_{1,2}\hdots \Theta_{N-1,N} M(\tau_1)\hdots M(\tau_N)\\\nonumber
& = \tau_9\Tr\left(\mathbbm1 + \int d\tau_1M(\tau_1) 
+ \int d\tau_1d\tau_2 M(\tau_1)M(\tau_2)\Theta_{12} + \dots \right)
\end{align}
where $\Theta_{12} = \Theta(\tau_1 - \tau_2)$.
It is simple to see (by working out the first few orders) 
that the potential part can be exponentiated.

\begin{align}\nonumber
e^{-S_{\p\Sigma}}& = \tau_9\left( V(T,T^\dagger) + 
\ap^2 \int d\tau_1d\tau_2 \Tr(\psi^\mu\p_\mu T_1 
\psi^\nu\p_\nu T_2^\dagger)sgn(\tau_{12}) \right.\\\nonumber
& - \ap^3\int d\tau_1d\tau_2d\tau_3 \Tr(\psi^\mu\p_\mu T_2 (T^\dagger T)_1 
\psi^\nu\p_\nu T_3^\dagger)(-\Theta_{32}\Theta_{21} - 
\Theta_{13}\Theta_{32}+\Theta_{21}\Theta_{13})\\\nonumber
& \left.- \ap^3\int d\tau_1d\tau_2d\tau_3 \Tr((TT^\dagger)_1\psi^\mu\p_\mu T_2 
\psi^\nu\p_\nu T_3^\dagger)(-\Theta_{31}\Theta_{12} + \Theta_{12}\Theta_{23} 
+ \Theta_{23}\Theta_{31}) \right)\\
& +  \mathcal O(\ap^4)
\end{align}
Where $V(T,T^\dagger)$ is given in (\ref{Potential}). The $sgn(\tau_{12})$
comes from a factor of $\Theta_{12} - \Theta_{21}$ and it reflects
the non-trivial ordering process involved in this path integral.
Putting one complex tachyon to zero, the two expressions at third order are the same 
($T^\dagger T = TT^\dagger = T\b T$) and
the various factors of $\Theta$ combine in pairs to give  1 if the $\tau$
are cyclically ordered or -1 if not.
In this case, one can then see the pattern and the kinetic
term exponentiate (together with the potential) to give the $\DD$ boundary insertion:
\begin{align*}
e^{-S_{\partial \Sigma}}|_{\DD} & = 2\tau_9 
e^{-\ap\int d\tau \left(TT^\dagger + \ap \psi^\mu\p_\mu T \frac{1}{\p_\tau}
\psi^\nu\p_\nu T^\dagger\right)}
\end{align*}
In the $\DDD$ case, the two terms at third order are 
different and the kinetic part
does not exponentiate any longer. We have:
\begin{align}
\psi^\mu\p_\mu T(T^\dagger T)\psi^\nu\p_\nu T^\dagger &= T_A\b T_A 
\psi^\mu\p_\mu T_A \psi^\nu\p_\nu \b T_A + T_B\b T_B \psi^\mu\p_\mu T_B 
\psi^\nu\p_\nu \b T_B\\\nonumber
& + T_A\b T_B \psi^\mu\p_\mu T_A \psi^\nu\p_\nu \b T_B + T_B\b T_A 
\psi^\mu\p_\mu T_B \psi^\nu\p_\nu \b T_A
\end{align}
\begin{align}
(T T^\dagger)\psi^\mu\p_\mu T \psi^\nu\p_\nu T^\dagger & = 
(T_A\b T_A + T_B\b T_B) 
(\psi^\mu\p_\mu T_A \psi^\nu\p_\nu\b T_A + \psi^\mu\p_\mu T_B 
\psi^\nu\p_\nu\b T_B)
\end{align}
These two expressions differ by various mixing terms between $T_A$ 
and $T_B$. So the path integral cannot be carried out in general.
Let us take the approximation where these mixing terms are ignored.
The task for finding an effective action is then
greatly simplified.
As we shall see, this approximate
effective action in closed form is still quite useful in
capturing some interesting physics.
This approximation physically signifies that
our action will only be valid when
1 tachyon is frozen while the other rolls.  We will see
in section \S4 how we can physically achieve such a situation using t-duality.

\subsection{The $\DDD$ Action}

Ignoring the mixing terms, the kinetic part will exponentiate
like before. Requiring the $U(2)$ symmetry in the path integral, 
we obtain, for the boundary insertion in the path integral
($T T^\dagger = T_A \b T_A + T_B \b T_B$) :
\begin{align}\label{approxBndy}
e^{-S_{\partial \Sigma}} & = \tau_9\left( 1 + 2 
e^{-\ap \int d\tau \left(TT^\dagger + \ap \psi^\mu\p_\mu T \frac{1}{\p_\tau}
\psi^\nu\p_\nu T^\dagger\right)}\right)
\end{align}
where the tachyon is now a SU(2) doublet.  
Although the mixing terms are important in some situations, 
they are absent in some dynamical situations. For example,
in the simple tachyon-rolling (that is, without forming defects),
all except one real tachyon should remain zero. In this case, the mixing 
terms are absent. Tachyon rolling that involves the formation 
of a codimension-2 (unstable) vortex involves only a single 
complex tachyon, so the mixing terms are absent again.
Mindful of this approximation, let us move on.

Consider a linear profile $T_i=u_iX^i$, $i=1,2,3,4$. 
This path integral (the usual disk action and the previous 
boundary insertion (\ref{approxBndy})) is gaussian and exactly solvable. 
One get (in flat spacetime), with $c_i=u_i^2$ :
\begin{align}\label{approxaction}
S_{(\DDD)_9} & = -\tau_9\int d^{10}x\left(1 + 2e^{-\la c_ix_i^2}
\prod_{i=1}^4\F\left(\frac{(\la)^2}{\pi}c_i\right)\right)
\end{align}
For a more general linear tachyon profile, $T=UX$ and $U$ is a 
$4\times 4$ matrix ($T$ and $X$ are 4-vectors for the 4 real tachyons).
Going to the diagonal basis for $U^\dagger U$ and let
$c_i$ be the eigenvalues of $U^\dagger U$, we have
$T T^\dagger = c_i x_i^2$. Again, we obtain (\ref{approxaction}).

We can generalize the above action to N D-branes and 1 anti-D-brane.  
The tachyon would be in the fundamental representation of SU(N) and 
we therefore will have N complex tachyons and 2N $\F$ in the action.  
The approximation is again that while 1 tachyon is rolling
all the other tachyons must be frozen.

\subsection{$U(2)\times U(1)$ covariant action}

As it is, our action (\ref{approxaction}) is not covariant under 
$U(2)\times U(1)$ since we have to choose a particular profile to be able
to solve the path integral. It would be interesting to covariantize 
this action, that is, replace the $c_i$ by gauge and Lorentz invariants
involving only first derivatives of $T$. 
We can then replace the derivatives by covariant derivatives. 
This turns out to be somewhat complicated, so we leave 
the details of the derivation in the appendix.
Here we give the main result. 
In the linear tachyon profile, we have $T=UX$, where $U$ is a 
$4 \times 4$ matrix. A general gauge and spacetime rotation can reduce the 
number of entries in $U$ from 16 to 6. Let us call them 
$u_i$, $i=1,2,..,6$. So we can express the 4 eigenvalues $c_i$ 
in terms of the 6 $u_i$.
Next we find that there are 6 independent gauge and Lorentz 
invariants that involve
only first derivatives of $T$. This set involve invariants up to
8 derivatives. Since we can write the $u_i$ in 
terms of the 6 invariants, the $c_i$ can also be written  
in terms of those invariants. This allows us to express the 
$\DDD$ action in terms of the 6 invariants.
The dependence of other invariants (with higher power of first
derivatives) in terms of a given basis 
is somewhat non-linear, so there is no obvious choice of basis.
A relatively simple choice of basis gives
\begin{align*}
4\pi\ap^2 c_1 = \X_+ + \sqrt{\Y_+} &\\
4\pi\ap^2 c_2 = \X_+ - \sqrt{\Y_+} &\\
4\pi\ap^2 c_3 = \X_- + \sqrt{\Y_-} &\\
4\pi\ap^2 c_4 = \X_- - \sqrt{\Y_-}
\end{align*}
\begin{align}\label{Invar}
  \X_{\pm} &\equiv 2\pi\ap^2\left(K_1 \pm \sqrt{K_3}\right)&
  \Y_{\pm} &\equiv (2\pi\ap^2)^2 \left(K_2 + K_4 \pm \sqrt{4K_2K_4 - K_5}\right)
\end{align}
\begin{align*}
K_1 & = \hf\p_\mu T_i\p^\mu\b T^i\\
K_2 & = \frac1{4}\left(\p_\mu T_i\p^\mu\b T^j \p_\nu T_j\p^\nu\b T^i - 
\epsilon_{ij}\epsilon_{mn}\p_\mu T_i \p^\mu\b T_m\p_\nu T_j\p^\nu\b T_n\right)\\
K_3 & =\frac1{4}\left(\p_\mu T_i\p^\mu T_j \p_\nu\b T^j\p^\nu\b T^i +
\sqrt{\epsilon_{ij}\epsilon_{kl}\epsilon_{mn}\epsilon_{ab} \p_\mu T_i\p^\mu T_k\p_\nu\b T_m\p^\nu\b T_a\p_\delta T_j\p^\delta T_l\p_\kappa\b T_n\p^\kappa\b T_b}\right)\\
K_4 & =\frac1{4}\left(\p_\mu T_i\p^\mu T_j \p_\nu\b T^j\p^\nu\b T^i -
\sqrt{\epsilon_{ij}\epsilon_{kl}\epsilon_{mn}\epsilon_{ab} \p_\mu T_i\p^\mu T_k\p_\nu\b T_m\p^\nu\b T_a\p_\delta T_j\p^\delta T_l\p_\kappa\b T_n\p^\kappa\b T_b}\right)\\
K_5 & =\p_\mu T_i \p^\mu T_j\p_\nu \b T^j\p^\nu T_k\p_\delta\b T^k\p^\delta T_m\p_\kappa\b T^m\p^\kappa\b T^i -\p_\mu T_i\p^\mu T_j\p_\nu\b T^j\p^\nu T_k\p_\delta\b T^k\p^\delta\b T^m\p_\kappa T_m\p^\kappa\b T^i
\end{align*}
where Greek indices refers to Lorentz indices and latin indices are 
SU(2) indices ($i=A,B$).
Now the action can be written in a covariant way:
\begin{align}\label{CovDDD}
S_{\DDD} =& -\tau_9\int d^{10}x\sqrt{-g}\left(1+  \right. \\
& \left. 2e^{-\la T^\dagger T}\FFFF\right) \nonumber
\end{align}
As is shown at the end of the appendix, the $U(1)$ and Lorentz 
invariants of the $\DD$
system arise naturally by setting $T_B = 0$. 

To T-dualize the above action such
 that we can move one brane away from the pair,
we need to covariantize further by restoring the gauge field.
The gauge group is $U(2)\times U(1) = SU(2)\times U(1)_-\times 
U(1)_+$ and, as usual, the tachyon does not couple to the $U(1)_+$. 
Therefore,

To restore $SU(2)\times U(1)_-$ we need to change the derivatives 
to covariant derivatives:
\begin{align*}
\p T \rightarrow D_\mu T & = \p_\mu T + i\left(A_\mu^a\frac{\sigma_a}{2} +\phi_\mu^-\right)T
\end{align*}

It turns out to be impossible to restore the gauge kinetic term simply. 
The simplest natural way would be a DBI-kind of prefactor 
\footnote{We would need to use the symmetric trace like was done in 
\cite{Tseytlin:1989rr}.} in front of
the whole action but one needs to make a difference between the gauge field
living on each of the two branes.
It must be that the gauge kinetic term mixes with the 
tachyon in a non-trivial manner.

Therefore, our $SU(2)\times U(1)$ invariant without the gauge field 
kinetic term is:

\begin{align}\label{GaugeAction}
S_{(\DDD)_9} & = -\tau_9\int d^{10}x \left(\sqrt{-g}\left(1+ 2e^{-\la TT^\dagger}\F^4\right)\right)
\end{align}

Where the arguments of $\F$ have been omitted and contain covariant 
derivatives.

\subsection{The RR Coupling}

Due to its topological nature, the RR action (\ref{generalCS}) should be 
exact for any number of branes and anti-branes \cite{Kennedy:1999nn,
Kraus:2000nj,Takayanagi:2000rz}.
For $\DDD$ action, the superconnection is given by:
\begin{align}
\mathcal F &= \left(\begin{array}{cc}
      F^{(1)} + iT^\dagger T & -i(DT)^\dagger\\
      -iD T & F^{(2)} + iTT^\dagger
    \end{array}\right)\nonumber
\end{align}
where $F^1$ and $T^\dagger T$ are $2\times 2$ matrices.  We use a 
shorthand notation $\mathcal T = T^\dagger T$ and $ t = TT^\dagger$.
\begin{align}
F^1 & = \left(\begin{array}{cc}
      F_{11} & F_{12}\\
      F_{21} & F_{22}
      \end{array}\right)\\
\mathcal T & \equiv T^\dagger T = \left(\begin{array}{cc}
      T_A\b T_A & T_A\b T_B\\
      T_B\b T_A & T_B\b T_B
      \end{array}\right)\nonumber\\\nonumber
t &\equiv TT^\dagger = T_A\b T_A + T_B\b T_B
\end{align}
For example, one can work out the coupling to $C_8$ for the case of 
2 D-branes and 1 anti-D-brane.  We just expand the exponential 
(\ref{generalCS}) keeping every 2-forms.  After 
some manipulations where the following identity is used, 
\begin{align*}
\Tr \left(\begin{array}{cc}
           -\mathcal T & \\
	   & -t
	   \end{array}\right)^n & = 
(-1)^n\Tr\left(\begin{array}{cc}
              \mathcal T^n & \\
	       & t^n
	       \end{array}\right) = (-1)^nt^{n-1}\Tr\left(\begin{array}{cc}
                                                  \mathcal T &\\
						  & t
						  \end{array}\right)
\end{align*}
we get($\lambda = \la$):
\begin{align}\label{CSform}\nonumber
\left.S_{RR}\right|_{C_8} & = \mu_9\int(-iC_8)\wedge 
\left(\frac{\lambda}{t}
(e^{-\lambda t} -1)DT\wedge DT^\dagger \right.\\\nonumber
& - \frac{\lambda}{t}(e^{-\lambda t}(\lambda +\frac{1}{t}) 
- \frac1{t})DT\mathcal{T}DT^\dagger + i\lambda (F_{11} + F_{22} - F^{(2)})\\ 
& \left.+ i\frac{\lambda}{t}
(e^{-\lambda t}-1)(Tr(\mathcal T F^{(1)}) - tF^{(2)})\right)
\end{align}

\section{T-duality of the $\DDD$ effective action}

Our action is only  valid when one tachyon rolls while the other 
is frozen. This scenario can be physically realized by moving 
one brane away, such that $T_B$ is no longer tachyonic. To do so, we 
have to consider $Dp$-branes where $p<9$.  This can be done 
using T-duality and we give the relevant formulas here.

The T-dual properties of the various fields in the
action are well known; the gauge fields in the T-dual directions
transform into the adjoint scalars, the metric and Kalb-Ramond field
mix, the string coupling scales and finally the field strength in the
T-dual direction gives rise to a commutator.
Being an open string scalar state,
the tachyon is inert under T-duality.  Under T-duality in directions 
labeled by uppercase Latin indices, (lowercase Latin indices
labeling unaffected directions on the brane), the fields transform as
\cite{Myers:1999ps}
\begin{align*}
  T &\to T,
  &A_a &\to A_a,
  &A_I &\to \frac{\Phi^I}{2\pi\ap},\\
  E_{\mu\nu} &\equiv g_{\mu\nu} + B_{\mu\nu},
  &e^{2\phi} &\to e^{2\phi}\det E^{IJ},
  &E_{IJ} &\to E^{IJ}\\
  E_{ab} &\to E_{ab}-E_{aI}E^{IJ}E_{Jb},
  &E_{aI} &\to E_{aK}E^{KI},
  &E_{Jb} &\to -E^{JK}E_{Kb},\\
  F_{aI} &\to \frac{D_a\Phi^I}{2\pi\ap},
  &F_{IJ} &\to \frac{i[\Phi^I,\Phi^J]}{4\pi^2\ap^2},
  &       &\\
\end{align*}
where $E^{IJ}$ is the matrix inverse to $E_{IJ}$ and it can therefore be used
to lower or raise indices. We used normal coordinates where the fields 
are independent of the coordinates we are 'dualing'. The non-commutative aspect
of the system appears in the T-dual of the field strength where we get
the commutator of $U(2)\times U(1)$
 matrix valued scalar fields. To treat this we follow 
\cite{Myers:1999ps} and introduce
\be
Q^I_J = \delta^I_J + i\la[\Phi^I,\Phi^K]E_{KJ}
\ee 
 The result of
T-dualing $9-p$ dimensions can be written most simply by defining the
pull-back in normal coordinates as:
\begin{align}
  &P[E_{ab}] \equiv E_{ab} + E_{aI}\p_b
  \Phi^I +
  \p_a\Phi^IE_{Ib}+
  \left(\p_a\Phi^IE_{IJ}\p_b\Phi^J\right),\\
  &P[E_{aI}] \equiv E_{aI} + \p_a\Phi^JE_{JI}\\
  &P[E_{Jb}] \equiv E_{Jb} + E_{JI}\p_b\Phi^I.
\end{align}
where the scalar $\Phi^I$ are:
\begin{align*}
\Phi^I & =\left(\begin{array}{cc}
           \Phi^{I\;1} & \\
	   & \Phi^{I\;2}
	   \end{array}\right)
= \left(\begin{array}{ccc}
        \Phi_{11}^{I} & \Phi_{12}^{I} & 0\\
	\Phi_{21}^{I} & \Phi_{22}^{I} & 0\\
	0 & 0 & \Phi^{I\;2}
	\end{array}\right)
\end{align*}
We can separate $\Phi^{I\;1}$ into a $SU(2)$ part $W^{I\;1}$ 
and a U(1) trace $\phi^{I\;1}$.
 The difference $\psi^I \equiv
\phi^{I\;1} - \Phi^{I\;2}$ is the scalar representing the
(DD - $\b{\text{D}}$)$_p$
%$(\text{DD}-\b \text{D})_p$ 
separation. The tachyon couples to $\varphi^I = \Phi^{I\;1} -
\Phi^{I\;2} = W^{I} + \psi^I$.
\begin{equation}\label{dualderivative}
D^I T = \p^I T + \frac{i}{\la}\varphi^I T
\end{equation}
In calculating the pull-back of any quantity
only the indices corresponding to the directions along the brane are
affected. After T-dualing the fields in (\ref{GaugeAction}) 
\footnote{It is important to realise that equation \ref{GaugeAction} does 
not have a DBI prefactor and T-duality will generate one naturally by mixing
the metric and the field strenght.  Like we
said previously, it is inconsistent to have such a global prefactor in front
of our action because it does not make the difference between the two D-branes.
For our sake, this is unimportant since
we only want to show how the tachyon gets a mass. One should not trust the 
DBI prefactor but the covariant devrivatives are still under control and give
 sensible results.}
as above and
performing manipulations similar to those in \cite{Myers:1999ps}, we
obtain the improved action for 2 D$_p$-branes and 1 anti-D$_p$-brane:
\begin{align}
  S_{(\DDD)_p} &= -\tau_p\int d^{p+1}x \sqrt{-\det[\G]} \left(1+ e^{-2\pi\ap 
TT^\dagger}\right.\\\nonumber
&\left.\FFFF\right)
\label{TDualAction}
\end{align}
%probably wrong need to do something about it.

where now the effective metric contains the spacetime metric
pulled-back to the brane worldvolume (and includes any non-zero NS-NS
B field) and has corrections from the non-commutativity of the coordinates.
\begin{align*}
  &\G_{ab} \equiv P[E_{ab} + E_{aI}(Q^{-1} -
  \delta)^{IJ}E_{Jb}]+2\pi\ap F_{ab}, 
\end{align*}
The covariant derivative dependence of $\X_{\pm}$ and $\Y_{\pm}$ in
(\ref{Invar}) leads to $\Phi$ dependence in the T-dual action.  Instead of
giving the complete expressions for $\X_{\pm}$ and $\Y_{\pm}$ (they are rather
long), we will show how to get all the essential features.  

The building blocks of all the invariants are 
$\p_{\mu}\b T_i G^{\mu\nu}\p_{\nu} T^i$ and two others contracted as tensor
($(TT)$ and $(\b T\b T)$). 
It is important to note that the SU(2) indices are not affected by
T-duality since they just refer to different scalars.  To get the T-duality of 
$\p_{\mu}\b T_i G^{\mu\nu}\p_{\nu} T^i$,
 we just need to compute how the inverse of the metric transforms and
how derivatives transform.
\begin{align}
[G_{\mu\nu}]^{-1} & = \left(\begin{array}{cc}
                G^{ba} & -G^{ba}P[E^{\prime}_{aI
}]\\
		P[E^{\prime}_{Jb}]G^{ba} & \left(E^{\prime}_{JI} - P[E^{\prime}_{Jb}]G^{ba}P[E^{\prime}_{aI}]\right) \end{array}\right)\\\nonumber
E^{\prime}_{aI} & \equiv E_{aJ} (Q^{-1})^J_I\\\nonumber
E^{\prime}_{Jb} & \equiv (Q^{-1})^K_J E_{Kb}\\
E^{\prime}_{JI} & \equiv E_{JL}(Q^{-1})^L_I \nonumber
\end{align}
We can now compute the transformation of our simplest
invariant ($K_1$) from the last equation.
It is completely similar to what was done in \cite{Jones:2002si} 
but we need to insert factors of $Q^{-1}$ in the pullbacks.  
\begin{align}
 \p_{\mu}\b T_i G^{\mu\nu}\p_{\nu} T^i&=
  Re\left[\begin{array}{r}
      \G_{1,2}^{ab}D_aT^\dagger D_b T 
      + \frac{1}{(2\pi\ap)^2}
      \left(E_{IJ}^{\prime} -P[E^\prime_{Ib}]\G^{ba} 
       P[E_{aJ}^{\prime}]\right)T^{\dagger}
       \varphi^{\dagger I}\varphi^JT\\
      -\frac i{2\pi\ap}\left(P[E^\prime_{Ia}]G^{ab}
      T^\dagger\varphi^{\dagger I}
       D_bT +
       G^{ab}P[E^\prime_{bI}] D_aT^\dagger\varphi^IT\right)
    \end{array}\right]
\end{align}
These expressions simplify considerably in Minkowski spacetime when
$B=0$ and $A^{1,2}=0$:
\begin{align}
  \G_{ab} = &\eta_{ab} + 
  (Q^{-1})_{IJ}(\partial_a\Phi^I\partial_b\Phi^J)\\
  \p_{\mu}\b T_i G^{\mu\nu}\p_{\nu} T^i \xrightarrow[B=0,\;A^{1,2}=0]{g=\eta}\; &
 \left[\G^{ab}\partial_aT^{\dagger} \partial_b T \right.\\
& -\frac{i}{2\pi\ap}\left((Q^{-1})_{IJ}\p_a\Phi^J
G^{ab}T^\dagger\varphi^{\dagger I}\p_bT +
G^{ab}\p_b\Phi^J(Q^{-1})_{JI}\p_aT^\dagger\varphi^I T\right)\nonumber\\\nonumber 
&\left. + \frac1{(2\pi\ap)^2}
    \left((Q^{-1})_{IJ} -  (Q)^{-1}_{IK}\p_b\Phi^K G^{ab}
  \p_a\Phi^M(Q)^{-1}_{MJ}\right) T^{\dagger}\varphi^{\dagger I}\varphi^J T\right]
\end{align}

The others invariants can be built the same way. 
The last expression is somewhat complicated because we kept every pieces.
It contains non-commutative effects 
which might be interesting to study. In this paper, we just want to 
set-up the formalism and
show explicitly how the tachyon gets a mass when we move one brane away.
Suppose we have 2 D$_8$-branes and 1 anti-D$_8$-brane that are transverse
to the $X_1$ direction, at $X_1=0$. Let us move one of the D-branes to
the position $\Phi_{11}$ while the remaining $\DD$-pair is held fixed at 
$\Phi^2 = \Phi_{22} = 0$.  The matrix $\Phi$ is therefore:
\begin{align*}
\Phi & \simeq \left(\begin{array}{ccc}
               \Phi_{11} & 0 & 0\\
	       0&0&0\\
	        0&0&0\\
		\end{array}\right)
\end{align*}
To simplify, we consider stationary branes : $\p\Phi = 0$. 
This yields a much simpler kinetic term: 
\begin{align}
\p_{\mu}\b T_i G^{\mu\nu}\p_{\nu} T^i & 
= \eta^{ab}\p_aT\p_bT^\dagger +\frac{1}{(\la)^2}T_A\Phi^*_{11}\Phi_{11} \b T_A
\end{align}
In this case, $T_A$ becomes massive and only $T_B$ remains tachyonic.

\section{Some Properties of the $\DDD$ action}

The formation of defects in a brane system has been studied using
K-theory arguments\cite{Witten:1998cd,Horava:1998jy}.
We expect that the tachyon condensation in the $\DDD$ system should 
produce only non-BPS defects. A simple example is a codimension-2 
defect, i.e., a vortex, produced by the annihilation of $\DD$ pair.
Such a vortex is unstable and it will eventually dissolve into
the remaining D-brane.

If we take the action (\ref{CovDDD}) and use it for the case where 
all the branes are sitting on top of each other,
we do not get any stable defects. As one
can see from table 1, we get the right tension for, the non-BPS, 
codimension 1 brane and for the vortex (codimension 2) but both of 
them are unstable due to the presence of the remaining tachyon. 
Instantons (codimension 4)
 would be stable but we do not get the right tension.
In fact, it is important to understand that the last two lines of table 1
do not represent any actual decay products of the tachyons.  
The mixing terms come in 
and it is impossible to neglect them any longer. In the following table the
first two lines represent possible unstable defects.  The last two lines,
on the other hand, give unphysical defects and are outside the region of 
validity of our approximation.

\begin{center}
Table 1: Resulting actions and tensions for linear tachyonic profile\\\vspace{0.5cm}
\begin{tabular}{|c|c|c|}
\hline
\parbox{3cm}{\begin{center}Number of rolling tachyon\end{center}} & Tension of solitons & Action after condensation\\
%rolling tachyon & $\left.\right.$ & $\left.\right.$ \\
%\hline
\hline
1 & $2\tau_9\sqrt{2\pi^2\ap} = \sqrt{2}\tau_8$ & $\int d^{10}x 
\biggl(\tau_9 +\sqrt{2}\tau_8 \delta(x_{10})e^{-\la T_i\b T^i}\F^3\biggr)$\\
\hline
2 & $2\tau_9(2\pi^2\ap) = \tau_7$ & $\int d^{10}x \biggl(\tau_9 +\tau_7\delta(x_{10},x_9) 
e^{-\la T_i\b T^i}\F^2\biggr)$\\
\hline
3 & $2\tau_9(2\pi^2\ap)^{\frac{3}{2}} = \frac{1}{\sqrt{2}}\tau_6$ & $\int d^{10}x 
\biggl(\tau_9 +\frac{1}{\sqrt{2}}\tau_6\delta(x_{10},x_9,x_8) e^{-\la T_i\b T^i}\F\biggr)$\\
\hline
4 & $2\tau_9(2\pi^2\ap)^2 = \hf\tau_5$ & $\int d^{10}x \biggl(\tau_9 
+\hf\tau_5\delta(x_{10},x_9,x_8,x_7)\biggr)$\\
\hline
\end{tabular}
\end{center}
Solitons can be obtained with a linear profile for the tachyon
$T_i = u_ix_i$ in the limit where $u_i$ goes to infinity.  
To get the tension 
of the solitons, we just have to put this profile into the action 
and integrate out the space over which the tachyon condenses 
(using the large x limit of $\F(x)$). 

The vortex is an interesting case since the action we get might be able to
describe the dissolution of a $\text{D}_{p-2}$ into a $\text{D}_p$.
This is the T-dual
analog of the recombination of strings \cite{Hashimoto:2003xz}.  
We believe that the following action, with addition of a gauge field coupling,
 might be able
to describe (at least in a qualitative level) the dissolution of branes 

\begin{align}\label{DpDpminus2}
S_{D_9D_7} & = -\tau_9\int d^{10}x\sqrt{-g} - \tau_7\int d^8x\sqrt{-g} e^{-\la T_B\b T_B}
\F(\mathcal X + \sqrt{\mathcal Y})\F(\mathcal X - \sqrt{\mathcal Y})
\end{align}
where the invariants are the same as in \cite{Jones:2002si}:
\begin{align*}
  \X &\equiv 2\pi\ap^2g^{\mu\nu}\partial_\mu T_B\partial_\nu\bar T_B,&
  \Y &\equiv\left(2\pi\ap^2\right)^2
  \Big(g^{\mu\nu}\partial_\mu T_B\partial_\nu T_B\Big)
  \Big(g^{\alpha\beta}\partial_\alpha\bar T_B\partial_\beta\bar T_B\Big),
\end{align*}
At first sight it might seems that the two parts of the action 
({\ref{DpDpminus2}) are completely
disconnected.  This is true as long as we only have the tachyon field.  
In order to describe the decay (dissolution), we 
need to restore the gauge fields. We will study this
issue elsewhere.
One can get stable vortices in our system if we take one brane away. 
In this case, only $T_A$ of the $\DD$ pair is present and the 
tachyon rolling simply gives the multi-vortex solitons of the $\DD$ system
\cite{Jones:2002si}. We can also include gauge flux inside the vortex.
\EPSFIGURE{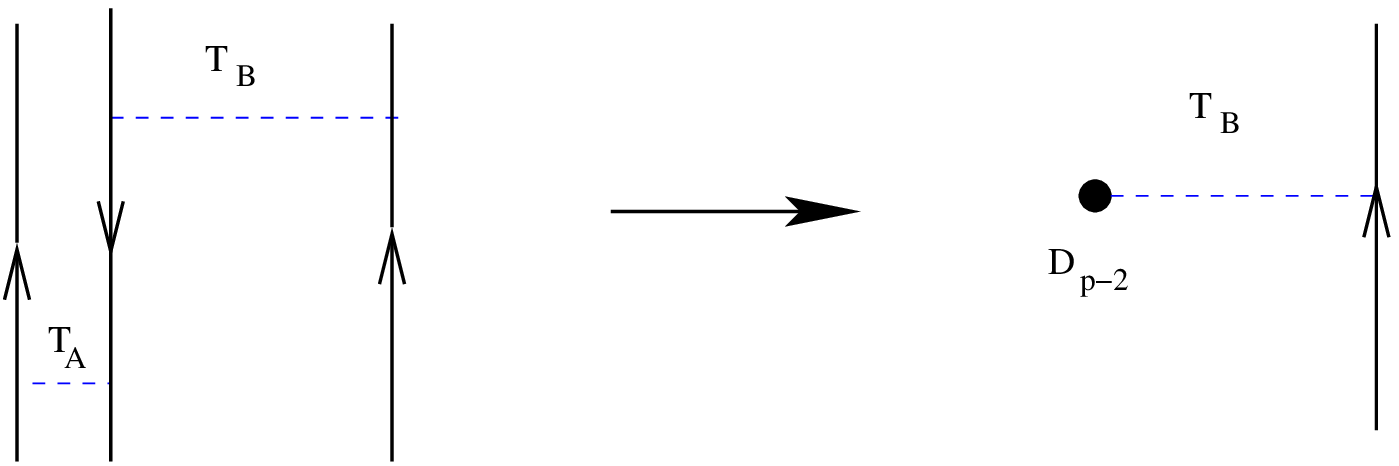}{A stable vortex is formed
if one brane is away from the $\DD$ pair such that $T_B$ is massive.
As the distance is reduced to the point where the $T_B$ becomes 
tachyonic, the vortex will dissolve.}

At the classical open string level, the brane separation is a modulus.
In the above case, $T_B$ has a real positive mass, so it is frozen 
at $T_B=0$, and the RR coupling with $D_{p-2}$-branes
simplifies considerably.
\begin{align}
\left.S_{RR}\right|_{C_8} & = -\frac{\tau_{p-2}g_s\lambda}{2\pi}
\int (-iC_{p-1})\wedge 
\left( -\frac{i}{\lambda}F_{22} \right.\\\nonumber
& \left.+ e^{-\lambda T_AT_A^\dagger}( DT_A\wedge DT_A^\dagger - 
\frac{i}{\lambda}(F_{22} - F^{(2)}))\right)
\end{align}
The coefficient in front comes from the relation between the 
tension of D$_p$ and D$_{p-2}$: 
$\mu_p\lambda = \tau_pg_s\lambda = \frac{\tau_{p-2}g_s}{2\pi}$.  
It gives the correct RR charge for a $D_{p-2}$-brane.

\section{Concluding Remarks}

In this paper, we attempt to generalize the $\DD$ effective 
action to that of the $\DDD$ effective action in boundary 
superstring field theory. Besides the non-abelian properties 
that are intrinsic in the $\DDD$ action, we find that the 
covariantization procedure also becomes quite complicated.
Taking the approximation where one tachyon is frozen while the
other rolls (which amount to neglecting all the cross-term between
the two tachyons),
we obtain a $\DDD$ action in closed form. We use it to construct 
the $D_pD_{p-2}$ action. The $D_pD_{p-2}$ system may provide 
an interesting brane inflationary model and will be  studied
in future work.

Another interesting application of this $\DDD$ system is to 
study how the system decay. It is clear that a $\DD$ pair in the
$\DDD$ system will annihilate, 
leaving behind a single D-brane. Where does the energy go?
In the pure $\DD$ system, the decay releases energy to some 
combination of defects, closed string modes and tachyon matter.
In the $\DDD$ system, the tachyons couple to $U(1)$ gauge fields
of the remaining brane. So, as the tachyon rolls, energy will 
also go to the open string modes which are absent in the pure
$\DD$ system. If we treat the $\DDD$ system 
as an inflationary model, energy that goes to open string modes
allow the heating of the universe at the end of inflation.
Since big bang cosmology puts strong bounds on the production 
of defects, closed string modes and tachyon matter, it will be 
important to determine the end products of tachyon rolling as 
a theoretical test of brane inflation.

\acknowledgments
We thank Horace Stoica and Saswat Sarangi for useful discussion.
This material is based
upon work supported by the National Science Foundation under Grant
No.~PHY-0098631.

\appendix\section{$SU(2)\times U(1)$ invariants}

The gauge group living on a system of coincident $M$ D-branes and 
$\bar M$ anti-D-branes is just $U(M)\times U(\bar M)$. 
The tachyon field transforms as a bifundamental. We are
interested in the case where $M=2$ and $\bar M=1$.  
The gauge symmetry is therefore $U(2) \times U(1)$. Since $T$ is 
charged only under a linear combination of the $U(1)$s, we need consider 
only $SU(2) \times U(1)$. We want to find a basis of 
gauge and Lorentz invariants made of only first derivative of $T$.

\subsection{Counting the number of invariants}

Start with $N=2n$ real tachyons in D spacetime dimensions, 
we have a general linear tachyon profile
$T_i=a_{ij}X_j$ with ND parameters $a_{ij}$. 
Let us consider $N<D$ only. We can always go to a 
N-dimensional subspace in D dimensions so we can reduce 
$a_{ij}$ to $N^2$ parameters ($i,j=1,2,...,N$). 
Spacetime rotation $O(N)$ has $N(N-1)/2$ generators and $U(n)$ has 
$n^2$ generators. We can further rotate (both spacetime and gauge) 
the linear profile to $K=N^2-N(N-1)/2 -n^2$ independent
parameters.  For the brane-antibrane case, n=1 and N=2, so there 
are $K=4-1-1=2$ parameters.
For the $U(2)$ case, we have $n=2$ and $N=4$, so there are $K=16-6-4=6$ 
parameters. We may choose them to be $u_1=a_{11}$, 
$u_2=a_{22}$, $u_3=a_{33}$, $u_4=a_{44}$ as well as 2 cross 
terms $u_5$ and $u_6$.
Therefore, based on the linear tachyon profile we expect that 6 invariants
will be needed even though we will have only 4 eigenvalues in our $\F$s.

The $U(1)$ case is easy to solve;
there are only 2 independent $U(1)$ and Lorentz invariants :
\begin{align*}
  \X &\equiv 2\pi\ap^2g^{\mu\nu}\p_\mu T\p_\nu\b T,&
  \Y &\equiv\left(2\pi\ap^2\right)^2
  \Big(g^{\mu\nu}\p_\mu T\p_\nu T\Big)
  \Big(g^{\alpha\beta}\p_\alpha\b T\p_\beta\b T\Big)
\end{align*}
That is, other invariants that involve only first derivatives can 
be written in terms of these two.

In the case of $U(2)$, there are 4 real tachyon fields,
or, equivalently, a complex doublet.
We will denote the two components of $T$ by the complex $T_1$ and 
$T_2$. 
The 2 basic $SU(2)$ invariant polynomials are the 
mesons and the baryons. For $U(1)$ invariance, each baryon 
must be accompanied by an antibaryon. Such a product can
be re-expressed as the antisymmetric sum of mesons :
\bas
M_{\mu\nu} & = & \p_\mu T^i\p_\nu\bar T_i\\
B^{\mu\nu} & = & \epsilon^{ij}\p^\mu T_i\p^\nu T_j\\
\tilde B_{\alpha\beta} & = & \epsilon^{ij}\p_\alpha\bar T_i\p_\beta\bar T_j\\
B^{\mu\nu}\tilde B_{\alpha\beta} & = & M^{[\mu}_\alpha M^{\nu]}_\beta 
\eas
where the Latin letters are $SU(2)$ indices
and greek letters are Lorentz indices.  Repeated indices are summed.  
Requiring $U(1)$ invariance allows us to write every gauge 
invariant polynomial in terms of mesons only. We would like to find the 
simplest 6 invariants such that all other invariants that involve 
only first derivatives can be expressed 
with them. We shall still use the baryons in our basis to
simplify the formula.

\subsection{Invariant Polynomials}

As we have learned in the $U(1)$ case, there are two different ways 
of contracting Lorentz indices: we can create a $SU(2)$ scalar by 
contracting $T$ with the corresponding $\b T$ or we can form a 
$SU(2)$ tensor by contracting $T$ 
with another $T$ (and doing the same for $\b T$).   
An invariant in which all the Lorentz contractions 
are done in such a way that $T$ is always contracting with a $\b T$ 
will be a "scalar type" term (with a subscript $s$).  
If all the contractions are  of the "tensor
type" we will use a subscript $t$.  When we reach terms with 
6 derivatives or more, there are invariants with both types of 
contractions; they will be labeled with a subscript $m$ for mixed.  
As we shall see, there are 2 mixed terms of 8 derivatives long and 
they will be differentiated with a prime.
Finally, the number of derivatives would be indicated as another 
subscript. Since this number is always even, the actual number of
derivative terms is twice the number in the subscript.
For example, the two terms in the $U(1)$ case are $M_{s1}$ and
$M_{t2}$. $M_{m4}$ means a mixed term
of 8 derivatives long
 and $B_{t4}$ means a baryon term contracted as a tensor of 8 
derivatives long. 

To simplify the notation further, 
we drop the derivative in front of $T$ since it
is always there, and we will denote Lorentz contraction by putting the two 
fields that are contracted into parenthesis.  $SU(2)$ contraction would be 
understood to be in cyclic order (first with the last, second with 
the third,...) for the mesons. The following brackets show the 
SU(2) contraction structure.
\be
M_{t4} = (\overbrace{T\underbrace{T)(\b T}\underbrace{\b T)(T}\underbrace{T)(\b T}\b T}) \nonumber
\ee
For baryons we will keep the $SU(2)$ clearly identified
with their $\epsilon$ tensor. 
%In the real field basis we have that
%\bas
%1 & \rightarrow & T_A = T_1 + iT_2\\
%2 & \rightarrow & T_B = T_3 + iT_4\\
%\eas
As examples :
\ba
M_{s1}  &= & \p_\mu T^i\p^\mu\bar T_i\rightarrow (T\b T) = 
(1\b 1) + (2 \b 2)\\
M_{t2}   &= &  \p_\mu T^i\p_\nu\b T_i\p^\mu T^j\p^\nu \b T_j  
\rightarrow \\\nonumber
(TT)(\b T\b T) &= &(11)(\b 1 \b 1) + (12)(\b 2 \b1) + (21)(\b 1 \b 2) 
+ (22)(\b 2 \b2)
\ea
Where $(1,2)$ refers to our 2 complex tachyons.  
Note that $(12)(\b 2 \b1) = (21)(\b 1 \b 2)$.
It is clear from this notation that we have six basic objects
$(11),(22),(1\b 1), (2\b2), (12),(1\b 2)$ and their complex conjugates. 
We like to show that there are 6 independent invariants.

Up to 8 derivatives there are at most 12 independent invariants
 We list them here:
\begin{center}
Table 2: Basic invariants up to 8 derivatives\\\vspace{0.5cm}
\begin{tabular}{|c|c|}\hline
name & polynomials\\
\hline\hline
$M_{s1}$ & $ \big(T\b T\big)  $ \\
\hline
$M_{s2}$ & $ \big(T\b T\big)\big(T\b T\big)$\\
\hline
$M_{t2}$ & $\big(TT\big)\big(\b T\b T\big)$ \\
\hline
$B_{s2}$ & $\epsilon_{ij}\epsilon_{mn}\big(T_i \b T_m\big)\big(T_j\b T_n\big)$\\
\hline
$M_{s3}$ & $\big(T\b T\big)\big(T\b T\big)\big(T\b T\big)$ \\
\hline
$M_{m3}$ & $\big(TT\big)\big(\b TT\big)\big(\b T\b T\big)$\\
\hline
$M_{s4}$& $\big(T\b T\big)\big(T\b T\big)\big(T\b T\big)\big(T\b T\big)$\\
\hline
$M_{t4}$& $(TT)(\b T\b T)(TT)(\b T\b T)$\\
\hline
$M_{m4}$&$(TT)(\b TT)(\b T\b T)(T\b T)$\\
\hline
$M_{m4}^{'}$&$(TT)(\b TT)(\b TT)(\b T\b T)$\\
\hline
$B_{t4}$&$\epsilon_{ij}\epsilon_{kl}\epsilon_{mn}\epsilon_{ab} (T_iT_k)(\b T_m\b T_a)(T_j T_l)(\b T_n\b T_b)$\\
\hline
$B_{m4}$&$\epsilon_{ij}\epsilon_{kl}\epsilon_{mn}\epsilon_{ab} (T_iT_k)(T_j\b T_a)(\b T_mT_l)(\b T_n\b T_b)$\\
\hline
\end{tabular}

\end{center}
Some of these terms can actually be expressed as a function of the others.
We give here a list of these relations.
First, there is an iterative relation between
$M_{sn}$, $M_{s(n-1)}$ and $M_{s(n-2)}$. From the first two 
invariants of this form we can get all the others.\footnote{The same 
relations exist for $M_{tn}$
but it will not be needed since we do not go beyond $M_{t4}$ which is only 
the second invariant that is totally contracted as a tensor.}
\be
M_{sn} = M_{s(n-1)}M_{s1} - \frac{(M_{s1}^2 - M_{s2})}{2}M_{s(n-2)}
\ee
Using this relation we learn that $M_{s3}$ and $M_{s4}$ are not 
independent. 
Furthermore, we already know that the baryons 
should be related to the mesons. Those relations are easy to find 
and they are:
\ba
B_{s2} & = &M_{s1}^2 - M_{s2}\\
B_{t4} &= &M_{t2}^2 - M_{t4}\\
B_{m4}& = &M_{m4} - 2M_{m4}^{'} + M_{t2}M_{s2}
\ea
Finally the last non trivial relation is:
\be
M_{m4}^{'}=M_{s1}M_{m3} - \frac{B_{s2}M_{t2}}{2}\\
\ee
This gives us a total of 6 relations among 12 invariants.
One can show that higher derivative invariants can be expressed in
terms of these 12, so we have a basis of 6 independent invariants, say:
$M_{s1}, M_{s2}, M_{t2}, M_{m3}, M_{t4}, M_{m4}$. As we shall see,
this basis is not the most suitable choice.

\subsection{Choice of a basis}

We will choose a basis which looks simple in 
the linear tachyon profile.  A general linear tachyon profile can 
be expressed in the real field basis by $T_i = a_{ij}X_j$ where $i$ 
is a gauge group indices 
($U(2)$ in our case and $i$ runs from 1 to 4) and $j$ is the 
spacetime index which
runs over the dimensionality of the space.  As pointed out earlier,
there is no lost of generality to restrict ourselves to a 4 dimensional 
subspace.  Therefore the $a_{ij}$ will represent a $4\times 4$ matrix 
and by a $SO(4)\times U(2)$ rotation we can eliminate 
10 off-diagonal elements of this matrix, leaving 6 non-zero matrix
elements in $a_{ij}$.
Note that, generically, it will not have the form ;
\[ T \ne \left( \begin{array}{cccc}
              u_1 & u_5 & 0 & 0\\
              0 & u_2 & 0 & 0\\
              0 & 0 & u_3 & u_6\\
              0 & 0 & 0 & u_4\\
	      \end{array}\right) X   \]
since further $U(1)$ rotations can render it completely diagonal with 
only 4 non-zero entries. In other words, it is 
impossible to get this linear profile by a $U(2) \times SO(4)$ 
transformation on a generic linear profile. 

For the rest of the analysis, we consider the following linear profile.
(We shall comment on other choices later.)
\[ T = \left( \begin{array}{cccc}
              u_1 & 0 & u_5 & 0\\
              0 & u_2 & 0 & u_6\\
              0 & 0 & u_3 & 0\\
              0 & 0 & 0 & u_4\\
	      \end{array}\right) X   \]
where $T$ and $X$ are 4 vectors. 
It is important to realize that for this basis all the 6 'pieces' 
($(11), (22), \dots$) are real.
We define 
\bas
(1\b 1)&= u_+,  (2\b 2) &=  v_+\\
(11) &= u_-,  (22) &= v_-\\
(1\b 2)&= w_+, (12) &=w_-\\
\eas
Now using that $(11) \rightarrow \p_{\mu}(T_1 +iT_2)\p^{\mu}(T_1 + iT_2)$ 
in the real field basis we can express the last expressions in terms of 
$u_1,u_2,\dots$:
\bas
u_{\pm} & = & u_1^2 + u_5^2 \pm (u_2^2 + u_6^2)\\
v_{\pm} & = & u_3^2 \pm u_4^2\\
w_{\pm} & = & u_5u_3 \pm u_6u_4\\
\eas
Using a matrix notation ($T=UX$) and, following \cite{Kraus:2000nj}, 
the functions 
that would appear in the lagrangian is $TT^\dagger = (XU)^TUX$. After doing 
the path integral, we will get 4 $\F$s with the variables being the 
eigenvalues of $U^TU$.
The eigenvalues of the system are ($U^TU$)
\bas
c_1&=& A + \frac{1}{2}\sqrt{C}\\
c_2&=& A - \frac{1}{2}\sqrt{C}\\
c_3&=& B + \frac{1}{2}\sqrt{D}\\
c_4&=& B - \frac{1}{2}\sqrt{D}\\
\eas
where
\bas
A&=&\frac{1}{2}(u_1^2 + u_3^2 +u_5^2)\\
B&=&\frac{1}{2}(u_2^2 +u_4^2 +u_6^2)\\
C&=&(u_1^2 - u_3^2)^2 + u_5^2(2u_1^2 +2u_3^2 +u_5^2)\\  
D&=&(u_2^2 - u_4^2)^2 + u_6^2(2u_2^2 + 2u_4^2 +u_6^2)\\
\eas
We can solve those eigenvalues in terms of 5 different terms in the 
basis (only
4 different combinations of 5 terms).
\begin{center}
Table 3: Invariants\\\vspace{0.5cm}
\begin{tabular}{|c|c|c|c|}
\hline
name & \# of derivatives &invariants& linear profile\\
\hline
$K_1$ & 2 &$\frac{1}{2}M_{s1}$ & $\frac{1}{2}(u_+ + v_+)$\\
\hline
$K_2$ & 4 &$\frac{1}{4}(M_{s2} - B_{s2})$ & $\frac{1}{4}( (u_+ - v_+)^2 + 4w_+^2)$\\
\hline
$K_3$ & 4 & $\frac{1}{4}(M_{t2} + \sqrt{B_{t4}})$ & $\frac{1}{4}(u_- + v_-)^2$\\
\hline 
$K_4$ & 4 & $\frac{1}{4}(M_{t2} - \sqrt{B_{t4}})$ & $\frac{1}{4}((u_- - v_-)^2 + 4w_-^2)$\\
\hline
$K_5$ & 8 & $(M_{m4}^{'} - M_{m4})$ & $(w_+(u_- -v_-) - w_-(u_+ - v_+))^2$\\
\hline
\end{tabular}
\end{center}

\bigskip

>From those basic invariant polynomials, we can solve for the $c$'s.  
Defining:
\begin{align*}
  \X_{\pm} &\equiv K_1 \pm \sqrt{K_3}&
  \Y_{\pm} &\equiv K_2 + K_4 \pm \sqrt{4K_2K_4 - K_5}
\end{align*}
from which we get
\begin{align*}
4\pi\ap^2 c_1 = \X_+ + \sqrt{\Y_+} &\\
4\pi\ap^2 c_2 = \X_+ - \sqrt{\Y_+} &\\
4\pi\ap^2 c_3 = \X_- + \sqrt{\Y_-} &\\
4\pi\ap^2 c_4 = \X_- - \sqrt{\Y_-}
\end{align*}

Another choice of the tachyon linear profile, for example, is
a symmetric matrix with 6 entries. It gives the same answer, suggesting
that our answer is independent of the choice of the basis. We have 
also considered a matrix which takes the form with a single 
eigenvalue, say $c_1$, while the remaining $3 \times 3$ matrix is 
in the Jordan canonical form with 5 $u_i$. In this case, we see 
rather easily that $c_1$ precisely reproduces the same function 
of the invariants given above.

Now we can write down our action (from BSFT with linear tachyon
profile and no gauge field) for the $\DDD$ in a general
covariant way. The action is proportional to:
\begin{align} \label{covariantSU}
\F (\X_+ + \sqrt{\Y_+})\F (\X_+ - \sqrt{\Y_+})\F (\X_- + \sqrt{\Y_-})
\F (\X_- - \sqrt{\Y_-})
\end{align}
It turns out that this expression does not have enough symmetry 
to cancel all
the square roots.  This could be a sign of non-local physics or simply a
limitation of our effective action.
It is easy to show that this action reduces to $\DD$ action if we set 
$T_2$ to 0.  In that case, the cross-terms and the v's
go to zero (no more baryons contributions
in the invariants) in the linear profile and there
is no longer any difference between $K_3$ and $K_4$.
In term of the invariants, we now have that $K_1^2 = K_2$ and $K_5=0$.
It is easy to see that $\Y_{\pm} = (K_1 \pm \sqrt{K_3})^2$ and we get:
\begin{align}
\X_+ + \sqrt{\Y_+} & \rightarrow 2(K_1 +\sqrt{K_3}) \\\nonumber
\X_+ - \sqrt{\Y_+} & \rightarrow  0\\\nonumber
\X_- + \sqrt{\Y_-} & \rightarrow 2(K_1 -\sqrt{K_3}) \\
\X_- - \sqrt{\Y_-} & \rightarrow 0 \nonumber
\end{align}
This is exactly the two invariants found in \cite{Jones:2002si}
and two of the $\F$'s just disappear ($\F(0) = 1$).

\providecommand{\href}[2]{#2}\begingroup\raggedright\endgroup


\begin{thebibliography}{10}

\bibitem{Sen:2002nu}
A.~Sen, {\it Rolling tachyon},  {\em JHEP} {\bf 04} (2002) 048,
  [\href{http://arXiv.org/abs/hep-th/0203211}{{\tt hep-th/0203211}}].

\bibitem{Sen:2002in}
A.~Sen, {\it Tachyon matter},  \href{http://arXiv.org/abs/hep-th/0203265}{{\tt
  hep-th/0203265}}.

\bibitem{Kraus:2000nj}
P.~Kraus and F.~Larsen, {\it Boundary string field theory of the {DD}-bar
  system},  {\em Phys. Rev.} {\bf D63} (2001) 106004,
  [\href{http://arXiv.org/abs/hep-th/0012198}{{\tt hep-th/0012198}}].

\bibitem{Takayanagi:2000rz}
T.~Takayanagi, S.~Terashima, and T.~Uesugi, {\it Brane-antibrane action from
  boundary string field theory},  {\em JHEP} {\bf 03} (2001) 019,
  [\href{http://arXiv.org/abs/hep-th/0012210}{{\tt hep-th/0012210}}].

\bibitem{Jones:2002si}
N.~T. Jones and S.~H.~H. Tye, {\it An improved brane anti-brane action from
  boundary superstring field theory and multi-vortex solutions},  {\em JHEP}
  {\bf 01} (2003) 012, [\href{http://arXiv.org/abs/hep-th/0211180}{{\tt
  hep-th/0211180}}].

\bibitem{Witten:1992qy}
E.~Witten, {\it On background independent open string field theory},  {\em
  Phys. Rev.} {\bf D46} (1992) 5467--5473,
  [\href{http://arXiv.org/abs/hep-th/9208027}{{\tt hep-th/9208027}}].

\bibitem{Gerasimov:2000zp}
A.~A. Gerasimov and S.~L. Shatashvili, {\it On exact tachyon potential in open
  string field theory},  {\em JHEP} {\bf 10} (2000) 034,
  [\href{http://arXiv.org/abs/hep-th/0009103}{{\tt hep-th/0009103}}].

\bibitem{Kutasov:2000aq}
D.~Kutasov, M.~Marino, and G.~W. Moore, {\it Remarks on tachyon condensation in
  superstring field theory},  \href{http://arXiv.org/abs/hep-th/0010108}{{\tt
  hep-th/0010108}}.

\bibitem{Tseytlin:1989rr}
A.~A. Tseytlin, {\it Sigma model approach to string theory},  {\em Int. J. Mod.
  Phys.} {\bf A4} (1989) 1257.

\bibitem{Marino:2001qc}
M.~Marino, {\it On the {BV} formulation of boundary superstring field theory},
  {\em JHEP} {\bf 06} (2001) 059,
  [\href{http://arXiv.org/abs/hep-th/0103089}{{\tt hep-th/0103089}}].

\bibitem{Niarchos:2001si}
V.~Niarchos and N.~Prezas, {\it Boundary superstring field theory},  {\em Nucl.
  Phys.} {\bf B619} (2001) 51--74,
  [\href{http://arXiv.org/abs/hep-th/0103102}{{\tt hep-th/0103102}}].

\bibitem{Quillen}
D.~Quillen, {\it Superconnections and the {C}hern character},  {\em Topology}
  {\bf 24(1)} (1985) 89--95.

\bibitem{Witten:1998cd}
E.~Witten, {\it {D}-branes and {K}-theory},  {\em JHEP} {\bf 12} (1998) 019,
  [\href{http://arXiv.org/abs/hep-th/9810188}{{\tt hep-th/9810188}}].

\bibitem{Marcus:1987cm}
N.~Marcus and A.~Sagnotti, {\it Group theory from `quarks' at the ends of
  strings},  {\em Phys. Lett.} {\bf B188} (1987) 58.

\bibitem{Witten:1982df}
E.~Witten, {\it Constraints on supersymmetry breaking},  {\em Nucl. Phys.} {\bf
  B202} (1982) 253.

\bibitem{Kennedy:1999nn}
C.~Kennedy and A.~Wilkins, {\it {R}amond-{R}amond couplings on brane-antibrane
  systems},  {\em Phys. Lett.} {\bf B464} (1999) 206--212,
  [\href{http://arXiv.org/abs/hep-th/9905195}{{\tt hep-th/9905195}}].

\bibitem{Sen:1999mg}
A.~Sen, {\it Non-{BPS} states and branes in string theory},
  \href{http://arXiv.org/abs/hep-th/9904207}{{\tt hep-th/9904207}}.

\bibitem{Tseytlin:1997cs}
A.~A. Tseytlin, {\it On non-abelian generalisation of the born-infeld action in
  string theory},  {\em Nucl. Phys.} {\bf B501} (1997) 41--52,
  [\href{http://arXiv.org/abs/hep-th/9701125}{{\tt hep-th/9701125}}].

\bibitem{Myers:1999ps}
R.~C. Myers, {\it Dielectric-branes},  {\em JHEP} {\bf 12} (1999) 022,
  [\href{http://arXiv.org/abs/hep-th/9910053}{{\tt hep-th/9910053}}].

\bibitem{Horava:1998jy}
P.~Horava, {\it Type {IIA} {D}-branes, {K}-theory, and matrix theory},  {\em
  Adv. Theor. Math. Phys.} {\bf 2} (1999) 1373--1404,
  [\href{http://arXiv.org/abs/hep-th/9812135}{{\tt hep-th/9812135}}].

\bibitem{Hashimoto:2003xz}
K.~Hashimoto and S.~Nagaoka, {\it Recombination of intersecting d-branes by
  local tachyon condensation},  \href{http://arXiv.org/abs/hep-th/0303204}{{\tt
  hep-th/0303204}}.

\end{thebibliography}
\end{document}